\documentclass[aps,prl,reprint,showpacs,nofootinbib]{revtex4-1}

\usepackage{graphicx,amsfonts,amssymb,amsmath,hyperref,bm}

\begin{document}

\title{Quantum loop expansion to high orders, extended Borel summation,\\
and comparison with exact results}

\author{Amna Noreen}
\email{Amna.Noreen@ntnu.no}
\author{K{\aa}re Olaussen}
\email{Kare.Olaussen@ntnu.no}

\affiliation{
Institutt for fysikk,\\
Norges Teknisk-Naturvitenskapelige Universitet,\\
N--7491 Trondheim, Norway
}

\begin{abstract}
We compare predictions of the quantum loop expansion
to (essentially) infinite orders with (essentially)
exact results in a simple quantum mechanical model.
We find that there are exponentially small corrections
to the loop expansion, which cannot be explained by any
obvious ``instanton'' type corrections.
It is not the mathematical occurence of exponential corrections,
but their seemingly lack of any physical origin, which we find surprising and puzzling.
\end{abstract}

\pacs{03.65.Sq, 02.30.Lt, 02.30.Mv}

\maketitle

\section{Introduction}

The Feynman path integral formulation\cite{FeynmanPathIntegral} is an
intuitive and powerful method of analyzing quantum systems.
The lowest order approximation can be understood in classical terms, with systematic
corrections available through a ``loop expansion'', which is
essentially an expansion in Planck's constant $\hbar$. The highlight of
such expansions is probably the recently completed tenth-order QED contribution
to the electron\cite{Electron_g-2_10thOrder} and muon\cite{Muon_g-2_10thOrder}
magnetic moments. The convergence of the resulting series is not yet
an acute issue for QED, but it is of practical interest for theories where the real
dimensionless expansion parameter is much greater, as f.i.~QCD. In fact,
it has been known since the argument of Dyson\cite{DysonNonconvergenceArgument}
that a power series in the fine structure constant cannot be convergent,
due to instability of QED if $\alpha= e^2/(4\pi\varepsilon_0\hbar c)$ changes sign.
An expansion in $\hbar$ is not quite the same, but a formal change of sign
of $\hbar$ also changes the sign of $\alpha$.
Hence, one should not expect more than asymptotic series, and hope
that they may be given well-defined and computable meaning through Borel
summation\cite{BorelSummation, OrszagBender, Weinberg}.

There are also genuine quantum phenomena, like tunneling processes, which
can be understood in quasi-classical terms. I.e., as classical
processes in imaginary time, often referred to as instanton corrections\cite{tHooftOnInstantons}.
They may lead to non-perturbative contributions which becomes
exponentially small as $\hbar\to 0$. In quantum field theory there may also
be ``renormalon'' contributions\cite{tHooftOnBorelSummation}
which obstructs a Borel summation. However, in simple models where the latter
phenomena do not occur one might think that the loop expansion provides a
complete description of the computed quantity. At least in principle.
At least we thought so.

In ordinary quantum mechanics an expansion in $\hbar$ should be
equivalent to a WKB expansion (although we are not aware
of any direct proofs of this). The latter seems much simpler
to carry out to high orders. The WKB expansion can be combined with a
quantization formula (\ref{exactquantization}) first written down
by Dunham\cite{DunhamHigherOrderWKBQuantizationFormula},
which to our knowledge has proven to be exact in all cases where
the result can be computed explicitly to all orders\cite{Bender_etal_NumerologicalAnalysis}.
One might get the impression that (\ref{exactquantization}) is always exact
(Dunham do not claim that). At least we thought so. Until we discovered otherwise.

We have analyzed the perhaps simplest model where the WKB result cannot be
computed explicitly to all orders. I.e., the eigenvalue problem 
\begin{equation}
   -\psi'' + x^4 \,\psi = E_N\,\psi,
   \label{AnharmonicOscillator}
\end{equation}
for large eigenvalue numbers $N$. Here the dimensionful quantity
$\hbar$ has been scaled out of the equation. We have used
$\delta_N = (N+\frac{1}{2})^{-2}$ as the real expansion parameter.
The WKB quantization formula for systems like this was derived to $12^{\text{th}}$ order
by Bender \emph{et.~al.}\cite{Bender_etal_NumerologicalAnalysis} (counting the standard
WKB approximation as $0^{\text{th}}$ order).
We have recently developed code for very high precision solutions
of Schr{\"o}dinger equations in one variable (and similar
ordinary differential equations)\cite{CPC2011:AsifAmnaKareIngjald, CPC2012:AmnaKare}.
In reference \onlinecite{CPC2011:AsifAmnaKareIngjald} we compared the $12^{\text{th}}$ order
approximation of (\ref{AnharmonicOscillator}) with our very-high-precision
numerical computations for eigenvalue number $n=50\,000$.
We found agreement to a relative accuracy of $5\times 10^{-67}$,
which is as expected of the WKB approximation for this value of $n$.
We interpreted this as a verification of the correctness of our numerical code, but it
does not constitute a very stringent test of the loop expansion itself.

For a more complete investigation of the latter we have extended
the WKB approximation to order $1\,704$.
This allows us to express the eigenvalues
of (\ref{AnharmonicOscillator}) as a series
\begin{equation}
    E_N = \text{const}\,\delta_N^{-2/3}\,\sum_{m\ge 0}\, t_m\,\delta_N^m,
    \label{InfiniteOrderWKBenergies}
\end{equation}
with $\delta_N = (N+\frac{1}{2})^{-2}$.
We have managed to construct a double integral representation
of the sum in (\ref{InfiniteOrderWKBenergies}),
with an expression for the integrand which is convergent over
the whole integration range. For small $\delta_N$ the result of
this representation is consistent with an ``optimal asymptotic approximation'' of
the sum, i.e.~summing the series up to (but not including)
the smallest absolute value. 

When comparing results of this approximation with numerical computations
to about $4\,000$ decimals accuracy, we find an intriguing discrepancy.
It vanishes exponentially fast as $\delta_N\to 0^+$, in a different manner
for even and odd eigenvalues.
But it is in both cases significantly larger than the expected uncertainty in
our evaluated WKB result.

The main lesson of our investigation
is that the asymptotic series from even very simple model calculations
may fail to provide results which are ``complete'', in the sense that
they have an accuracy of the same magnitude as the accuracy
inferred from the optimal asymptotic approximation or
a Borel summation of the series.

In the remainder of this letter we present some details of our
computations and results.

\section{WKB and Dunham formulas}

For the WKB approximation we formally
change equation~(\ref{AnharmonicOscillator}) to
\(
  -\epsilon^2 \psi'' + x^4 \,\psi = E\,\psi
\),
write $\psi = \text{e}^{S}$, and expand 
\(
  S = \epsilon^{-1}\sum_{n\ge 0} \epsilon^n S_n
\)
to obtain (with $Q = V-E$)
\begin{subequations}
    \label{WKBrecursion}
    \begin{align}
    &{S'_0}^2 = -Q,\label{WKBinitialization}\\
    &S''_{n-1} + \sum_{j=0}^{n} S'_{j} S'_{n-j} = 0\label{WKBrecursionStep},
    \end{align}
\end{subequations}
which can be solved recursively.
The Dunham quantization formula reads
\begin{equation}
  \frac{1}{2\text{i}}\oint S'(z) \text{d}z = N\pi,
  \label{exactquantization}   
\end{equation}
where the integral encircles a branch cut between the two classical
turning points of (\ref{AnharmonicOscillator}) at
$z = \pm E^{1/4}$. All odd terms beyond $n=1$ in the
series for $S$
can be written as the derivative of a
function which is single-valued around the integration
contour\cite{DunhamHigherOrderWKBQuantizationFormula, Bender_etal_NumerologicalAnalysis},
and hence does not contribute to the quantization condition~(\ref{exactquantization}).
The even terms may also be simplified by adding derivatives of single-valued functions.

\section{Explicit computations}

For our case, where $V=x^4$, the integral in equation~(\ref{exactquantization}) can be
reduced to a sum of integrals of the form
\begin{subequations}
    \label{QuantizationIntegrals}
    \begin{align}
          I^{\text{(e)}}_{k}  &= \frac{1}{2\text{i}}  \oint \frac{1}{\left(z^4 - E\right)^{k+1/2}}\,dz\label{Ie_k}\\
          &= (-1)^{k} \frac{1}{2}\,\mathrm{B}(\textstyle{\frac{1}{4}},\textstyle{\frac{1}{2}-k})\,E^{-1/4 -k},
          \quad\text{for $k=6\ell-1$},\nonumber
          \\
          I^{\text{(o)}}_{k}  &= \frac{1}{2\text{i}} \oint \frac{z^2}{\left(z^4 - E\right)^{k+1/2}}\,dz\label{Io_k}\\
          &= (-1)^{k} \frac{1}{2}\, \mathrm{B}({\textstyle \frac{3}{4}},{\textstyle \frac{1}{2}}-k)\,E^{1/4-k},
          \quad\text{for $k=6\ell+2$}.\nonumber
    \end{align}
\end{subequations}


\noindent
Here $\mathrm{B}(a, b)
$ is the Beta function.
These integrals
must be multiplied by factors
$(-1)^{\ell}\,p_{\ell}^{\text{(e)}} E^{3\ell}$ and
$(-1)^{\ell}\,p_{\ell}^{\text{(o)}} E^{3\ell+1}$
respectively, where the $p_{\ell}$'s are positive rational numbers found by solving equation~(\ref{WKBrecursionStep}).
F.i.,~$p^{(\text{o})}_0 = \frac{1}{2}$, $p^{(\text{e})}_1 = \frac{77}{1\,768}$,
$p^{(\text{o})}_1 = \frac{61\,061}{62\,928}$.
Inserted into (\ref{exactquantization}) we obtain
\begin{align}
    &\frac{1}{3\epsilon} \mathrm{B}({\textstyle \frac{1}{4},\frac{1}{2}})\,E^{3/4} -
    \frac{\epsilon}{16} \mathrm{B}({\textstyle \frac{3}{4},\frac{1}{2}})\,E^{-3/4}\nonumber\\
    &+ \sum_{\ell=1}^{\infty}  (-1)^{\ell} \mathrm{B}({\textstyle \frac{1}{4},\frac{1}{2}}) 
    q^{\text{(e)}}_{\ell}\,\epsilon^{4\ell-1} E^{-(12\ell-3 )/4}  
    \label{InfiniteOrderQuantization}\\ &+  \sum_{\ell=1}^\infty(-1)^{\ell} 
    \mathrm{B}({\textstyle \frac{3}{4},\frac{1}{2}})
    q^{\text{(o)}}_{\ell} \epsilon^{4\ell+1}\,E^{-(12\ell+3)/4}
    = \left(N+{\textstyle \frac{1}{2}}\right)\pi,\nonumber
\end{align}
where $q^{\text{(e)}}_{\ell}$ and $q^{\text{(o)}}_{\ell}$ are positive rational numbers.
By introducing
\begin{equation}
  \varepsilon \equiv \left[{3\pi}/{\mathrm{B}({\textstyle \frac{1}{4},\frac{1}{2}})}\right]^2 \epsilon^2\,E^{-3/2},
  \label{RescaledEnergy}
\end{equation}
which is a small quantity for large quantum numbers $N$,
we can rewrite equation (\ref{InfiniteOrderQuantization}) as
{\footnotesize
\begin{equation}
     \varepsilon  = \frac{1}{(N + \frac{1}{2})^2}\left(\sum^\infty_{\ell=0}  (-1)^{\ell+1} 
       \left[ r_{2\ell}\, \varepsilon^{2\ell} + r_{2\ell+1}\, \varepsilon^{2\ell+1} \right]\right)^2
     \label{InfiniteOrderQuantization2}
\end{equation}
}

\noindent
with new coefficients $r_0=-1$, $r_1={1}/{12\pi}$, and all other $r_{m}$ positive numbers
growing like $m!^2 a_r^m (m+1)^\nu$ for large $m$. 
We have computed these coefficients up to $m=852$  (selected at random by a license server failure).
Empirically they fit the cited behaviour quite well, with  $a_r \approx 0.202\,641\,423\,4$
and $\nu = -\frac{5}{2}$, and systematic higher order correction as a series in $m^{-1}$.
Obviously the sum in equation~(\ref{InfiniteOrderQuantization2}) has zero radius of convergence,
but it can be turned into a well defined integral by (essentially) a twice iterated
Borel summation.

Here we will first invert equation~(\ref{InfiniteOrderQuantization2}) to an explicit expression
for the eigenvalues $E_N$. By introducing $\delta_N \equiv (N+\frac{1}{2})^{-2}$ we can first express
$\varepsilon \equiv \varepsilon_N$ as a series in $\delta_N$,
\begin{equation}
   \varepsilon_N = \delta_N + \sum_{m\ge 2} s_m \delta^m_N,
\end{equation}
where the coefficients $s_m$ can be computed recursively. Their explicit analytic
expressions are rational polynomials in $\mathrm{B}(\frac{1}{4},\frac{1}{2})$ and $\pi^{-1}$,
which soon become too complicated for practical use. F.i.,
\begin{equation*}
{\textstyle
  s_2= -\frac{1}{6\pi},\quad s_3 = \frac{11}{20\,736\,\pi^4}\,\textrm{B}(\frac{1}{4},\frac{1}{2})
  + \frac{5}{144\,\pi^2}.
}
\end{equation*} 
Instead we have computed $s_m$ numerically to about $3\,800$ decimals accuracy up to $m=852$.
Finally we use~(\ref{RescaledEnergy}) to express $E_N$ by a series in $\delta_N$. With the
formal expansion parameter $\epsilon=1$,
\begin{align}
    E_N &= \left[{3\pi}/{\mathrm{B}({\textstyle \frac{1}{4},\frac{1}{2}})}\right]^{4/3}\, \delta_N^{-2/3}\,
    \Big{(}1 + \sum_{m\ge1} s_m \delta_N^m \Big{)}^{-2/3} \nonumber\\
    &=  \left[{3\pi}/{\mathrm{B}({\textstyle \frac{1}{4},\frac{1}{2}})}\right]^{4/3}\,  \delta_N^{-2/3}
    \,\Big{(} 1 + \sum_{m\ge1}  t_m \,\delta_N^m \Big{)}.
    \label{EigenvalueSeries}
\end{align}

\begin{figure}
\begin{center}
\includegraphics[clip, trim= 10ex 6.5ex 10ex 5ex,width=0.5\textwidth]{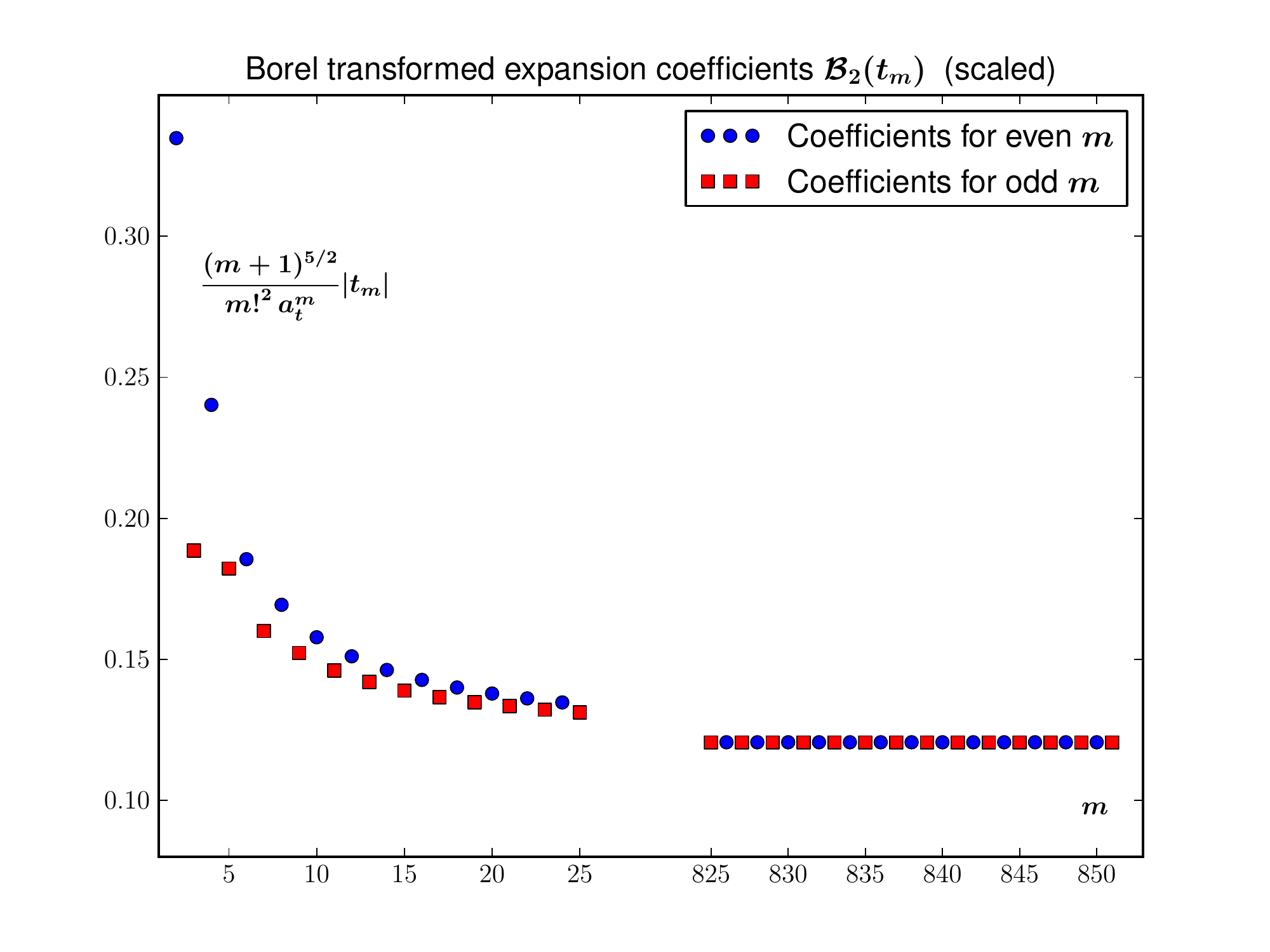}
\end{center}
\caption{\label{BtCoeffs}Rescaled form of the expansion coefficients $t_m$ in 
equation (\ref{EigenvalueSeries}) for $m=0,\ldots,852$.
The index $\nu=\frac{5}{2}$ is chosen as the simplest rational number close to
the best fit,  after which we find $a_t \approx 0.202\,641\,423\,4$ as the best fit to a sequence approaching
a constant absolute value for large $m$.
}
\end{figure}

The first few terms are
\begin{equation*}
{\textstyle
   t_1 =   \frac{1}{9\pi},\quad 
   t_2 = -\frac{5}{648\,\pi^2}- \frac{11}{31\,104\,\pi^4}\, \mathrm{B}(\frac{1}{4},\frac{1}{2}).
}
\end{equation*}
The coefficients $s_m$ and $t_m$ grow in magnitude in essentially the same manner as $r_m$,
i.e.~$\vert t_m \vert \sim m!^2 a^m_t (m+1)^{\nu}$ with $a_t = a_r$ and $\nu=-\frac{5}{2}$
(cf. figure~\ref{BtCoeffs}).
The even and odd sequences behave slightly different.
Beyond $t_0=1$ the coefficients $t_{2\ell}$ and $t_{2\ell+1}$ have sign $(-1)^{\ell+1}$. 

\section{Extended Borel summation}

To make sense of the sum in (\ref{EigenvalueSeries}) we use the formula
\begin{align}
    m!^2 =\alpha^{2(m+1)}
    \int_0^{\infty} dx\,x^m\,\text{e}^{-\alpha x}\,
    \int_0^{\infty} dy\,y^m\,\text{e}^{-\alpha y}
\end{align}
where $\alpha = \text{e}^{\text{i}\phi}$, with $-\frac{1}{2}\pi< \phi < \frac{1}{2}\pi$.
Interchange of summation and integration gives a Borel sum
($z\equiv x y a_t \delta$),
\begin{equation}
   t(\delta) = \int_0^\infty dx\,\text{e}^{-\alpha x}\,\int_0^\infty dy\,\text{e}^{-\alpha y}\,
   \sum_{m=0}^\infty\,\tilde{t}_{m}\,z^m, 
   \label{BorelResummation}
\end{equation}
with
\(
\tilde{t}_m = {\alpha^{2(m+1)}}\,t_m/({{m!}^2\,a_t^m})
\).
Our computations indicate that the sum
\(
  \tilde{t}(z) \equiv \sum_{m=0}^\infty \tilde{t}_m\, z^m
\)
converges\footnote{We find the equivalent form
$t(\delta) = 2\int_0^\infty d\xi\,\xi^m
\,K_0(2\alpha\sqrt{\xi})\,\tilde{t}(\xi a_t \delta)$ to be less
convenient for further manipulations.} for $\vert z \vert < 1$.
For $\alpha=1$ the function $\tilde{t}(z)$ has singularities where $z^2=-1$, 
with the singular parts behaving like $(1+z^2)^{3/2}$.
In terms of the variable ${z^2}/{(1+z^2)}$ the points $z^2 = -1$ are mapped to $\infty$, 
and the full integration range to the interval $[0,1]$. 
We tried this substitution in the hope that the resulting sums for $\tilde{t}(z)$ would converge over the
full integration range, but discovered additional singularites for $z^2 \approx 4$.

Hence, to avoid integrating through a singularity, 
one must introduce the phase $\alpha$ (or equivalently integrate along a
different direction in the complex plane). A convenient choice is $\alpha=\text{e}^{\text{i}\pi/8}$,
or its complex conjugate. Actually, to assure a real result after analytic continuation of 
$\tilde{t}(z)$, one must take the average of these two choices.
This amounts to taking the real part of the integral~(\ref{BorelResummation}).

After our choice of $\alpha$ we separate $\tilde{t}(z)$ into four (infinite) sums,
\(
    \tilde{t}(z) = \sum_{p=0}^3  z^p\, \sum_{\ell\ge 0}  \tilde{t}_{4\ell+p} z^{4\ell}
\).
The function defined by each infinite sum is singular at $z^4=-1$, $z^4\approx -16$, and probably
at infinitely many more points on the negative real $z^4$-axis. Next we rewrite
\begin{equation}
      \sum_{\ell\ge0}  \tilde{t}_{4\ell+p}\, z^{4\ell} = 
      \sum_{\ell\ge0} \hat{t}_{4\ell+p}\,\left({\textstyle \frac{z^4}{1+z^4}}\right)^\ell \equiv \hat{t}^{(p)}(z),
      \label{PadeResummation}
\end{equation}
and use the computed coefficients $\tilde{t}_{4\ell+p}$
to find equally many coefficients $\hat{t}_{4\ell+p}$. A ratio test on the coefficients 
$\hat{t}^{(p)}_{\ell} \equiv \hat{t}_{4\ell+p}$ indicate that the second sum in (\ref{PadeResummation}) converges
for $\vert z^4/(1+z^4) \vert < 1$ for all four values of $p$. This provides an expression,
\begin{equation}
   t(\delta) = \text{Re} \int_0^\infty dx\,\text{e}^{-\alpha x}\,\int_0^\infty dy\,\text{e}^{-\alpha y}\,
   \sum_{p=0}^3 z^p\, \hat{t}^{(p)}(z),
   \label{BorelSumRepresentation}
\end{equation}
where each $t^{(p)}(z)$ is computable by a convergent power series
in $u\equiv z^4/(1+z^4)$ over the full integration range. We have
computed $\hat{t}_\ell^{(p)}$ for $\ell \le 212$.

\begin{figure}
\begin{center}
\includegraphics[clip, trim= 10ex 6.5ex 10ex 1.75ex,width=0.5\textwidth]{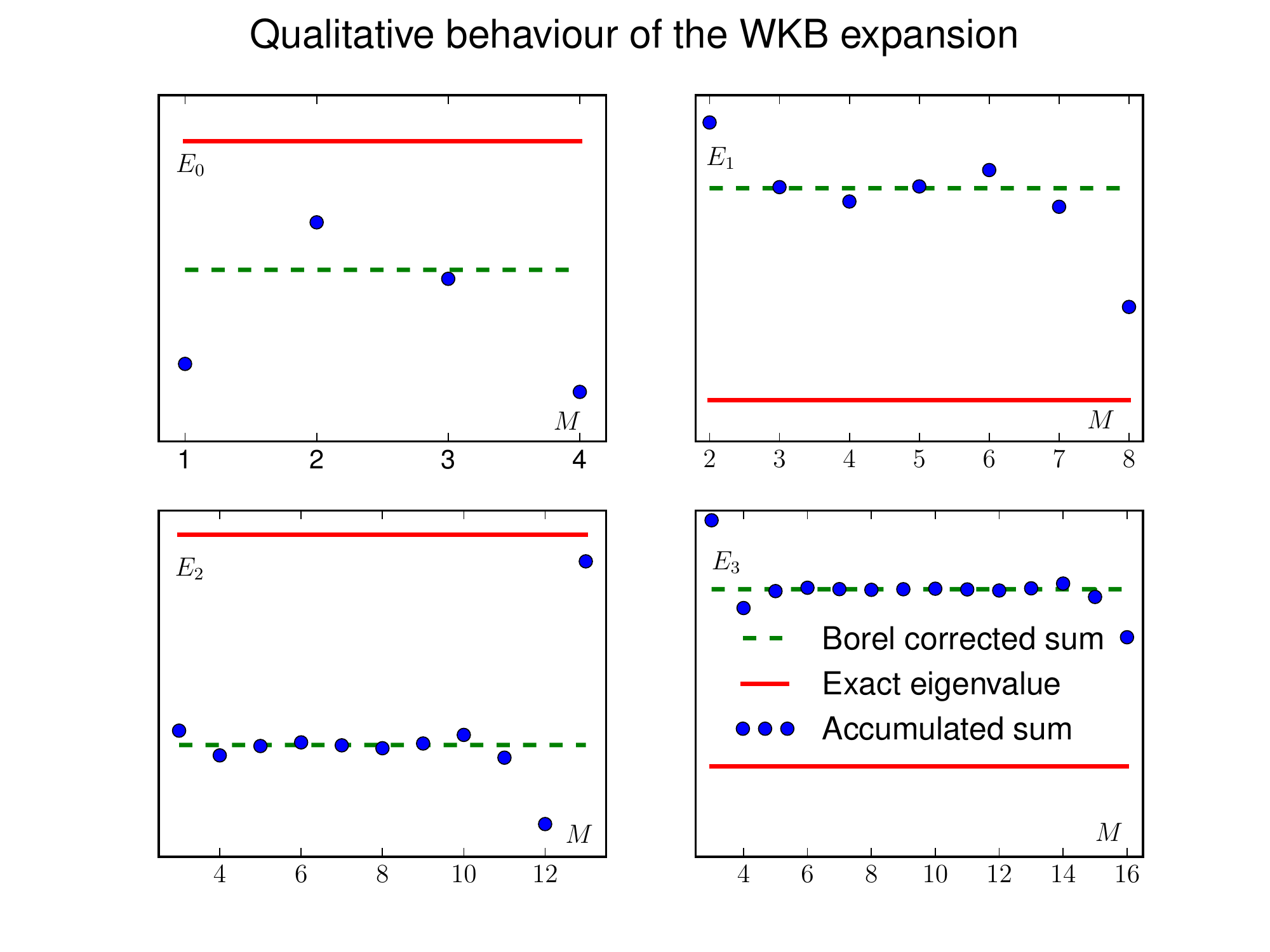}
\end{center}
\caption{\label{WKB_behaviour}
Behaviour of the WKB expansion of the few lowest
eigenvalues $E_N$. The points show the sum $\sum_{m=0}^{M-1} t_m\,\delta^m$
in equation~(\ref{BorelCorrectedExpansion}). The (green) dashed lines show the full expression
$t(\delta)$; it is independent of $M$ to the expected numerical accuracy
of the correction term $t_{\text{corr}}^{(M)}(\delta)$. The red lines show the exact eigenvalue,
evaluated numerically by our very-high-precision routine. Obviously the
Borel corrected WKB series does not converge towards the exact eigenvalues.
}
\end{figure}

The representation (\ref{BorelSumRepresentation}) is not optimal for evaluating
$t(\delta)$ for small $\delta$. Instead write
$\text{e}^{-\alpha x} = -\alpha^* \frac{d}{dx} \text{e}^{-\alpha x}$
and perform a partial integration in $x$. Repeating this $M$ times
regenerates the $M$ first terms of the series in (\ref{EigenvalueSeries}),
with the remaining coefficients available to construct a correction term,
\begin{equation}
     t(\delta) = \sum_{m=0}^{{ M}-1 }t_m\,\delta^m + t^{(M)}_{\text{corr}}(\delta).
     \label{BorelCorrectedExpansion}
\end{equation}
Here $t^{(M)}_{\text{corr}}(\delta)$ is an integral similar to $t(\delta)\equiv t^{(0)}_{\text{corr}}(\delta)$.
It must be computed numerically, but the integral is proportional to the exactly
known prefactor $t_M\,\delta^M$ which may be small. A consistency check is
that $t(\delta)$ should be independent of $M$, at least for a range of
$M$-values around the minimum of $\big{|} t^{(M)}_{\text{corr}}(\delta) \big{|}$.
As shown in figure~\ref{WKB_behaviour} for the four lowest eigenvalues,
the representation~(\ref{BorelCorrectedExpansion}) provides results
which are independent of $M$ within the accuracy of the numerical
integration. {These results are significantly different from the
exact eigenvalues}.

\begin{figure}
\begin{center}
\includegraphics[clip, trim= 10ex 6.5ex 10ex 1.75ex,width=0.5\textwidth]{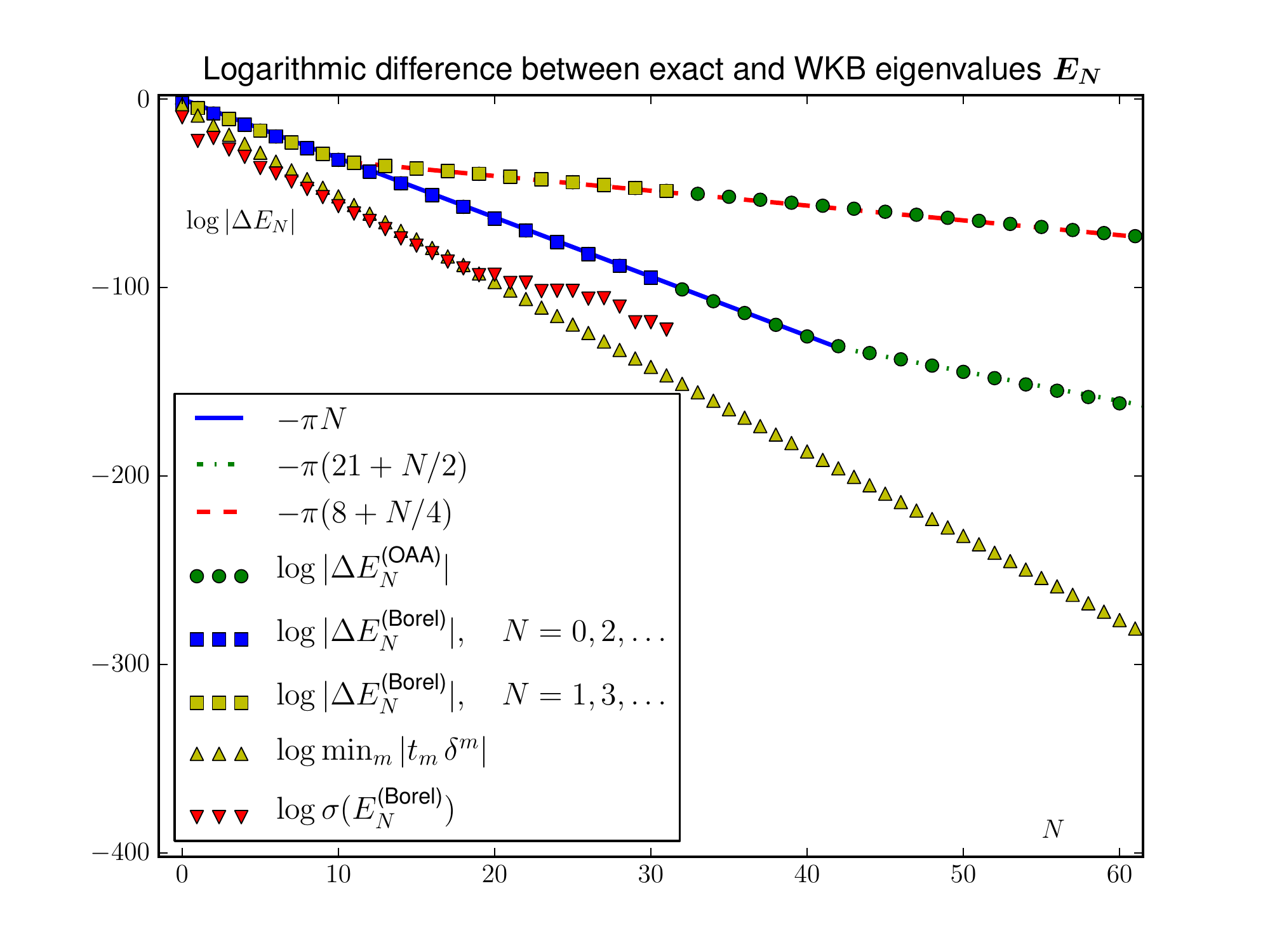}
\end{center}
\caption{\label{logDeltaEn2}Difference between the exact eigenvalues $E_{N,\text{exact}}$
(computed numerically to very high precision) and the WKB eigenvalues $E_{N,\text{WKB}}$,
computed using either the optimal asymptotic approximation (OAA) or adding the correction
integral from Borel summation (Borel). The results of these two methods cannot be distinguished
in the figure when $N\ge 1$.
The later is found from equation~(\ref{RescaledEnergy}),
with $t(\delta)$ computed from equation~(\ref{BorelCorrectedExpansion}) for a range of $M$-values around
$2.2 N$. The result varies little with $M$, as indicated by the plotted standard deviation $\sigma(E_N)$.
Hence, the difference between $E_{N,\text{exact}}$ and $E_{N,\text{WKB}}$ is much larger than the
uncertainty in $E_{N,\text{WKB}}$ due to numerical evaluation of the integral for
$t_{\text{corr}}^{(M)}(\delta)$, although exponentially small as function of $N$.
The correction terms look quite simple, with an interesting difference between the even and odd eigenvalues.
}
\end{figure}

When investigating a larger range of $N$ we find
the representation (\ref{BorelCorrectedExpansion}) to be consistent
with the optimal asymptotic approximation, which is much faster and easier to
evaluate. For $N \ge 1$ the results of the two methods cannot be
distinguished when compared with the distance to the exact eigenvalue.
Also, the numerical uncertainty in (\ref{BorelCorrectedExpansion}) is
comparable to the smallest term in (\ref{EigenvalueSeries}) (and
eventually worse as $N$ increases). This is shown in figure~\ref{logDeltaEn2},
where we plot $\log\vert E_{N,\text{exact}} - E_{N,\text{WKB}}\vert$ as
function of $N$, together with the expected uncertainties in the
evaluated values of $E_{N,\text{WKB}}$. 

For low $N$ we find empirically (to exponential accuracy) that 
\begin{equation}
   E_{N,\text{exact}} - E_{N,\text{WKB}} \approx (-1)^N\,\text{e}^{-\pi N},
\end{equation}
but, intriguingly, this behaviour is overtaken by larger error terms
when $N \ge 11$ for odd $N$, and $N \ge 44$ for even $N$. We still
find $E_{2N,\text{exact}} - E_{2N,\text{WKB}}$ to be positive, but now
the sign of $E_{2N+1,\text{exact}} - E_{2N+1,\text{WKB}}$ is $(-1)^N$.

\section{Concluding remarks}

The formula~(\ref{exactquantization}), with $S$ computed by WKB to all orders,
does not always provide exact eigenvalues, but at best the complete contribution
of the WKB approximation. With hindsight it is clear that there may be
exponential small corrections: WKB does not predict
backscattering of a forward propagating wave to any
order of approximation when $E-V > 0$. This is
generally known to occur in exact calculations.
Double backscattering is likely to contribute an exponentially
small correction to the left hand side of (\ref{exactquantization}).
Also in asymptotic analysis a second exponential behavior is said to
emerge when Stokes lines are crossed, but this statement alone is
unhelpful for actually computing any exponentially small correction.

\section*{Acknowledgment}

We thank A.~Mushtaq and I.~{\O}verb{\o} for useful discussions.
This work was supported in part by the Higher Education
Commission of Pakistan (HEC).

\end{document}